\documentclass[12pt]{article}
\input{tcilatex}

\begin{document}

\begin{center}
{\LARGE \textbf{Generalized Quantum Theory:\\[0.3cm]
Overview and Latest Developments}} \vspace{1.2cm}

{\large \textbf{Thomas Filk${}^{1,2}$ and Hartmann R\"omer${}^1$}} \vspace{%
0.5cm}

{\normalsize ${}^1$ Department of Physics, University of  Freiburg,}\\[-0.1cm%
]
{\normalsize Hermann-Herder-Str.~3, D-79104 Freiburg}\\[0.1cm]
{\normalsize ${}^2$ Parmenides Foundation for the Study of Thinking, Munich}%
\\[0.5cm]
{\normalsize Part of this project was financed by the Fetzer-Franklin Fund}
\end{center}

\thispagestyle{empty} \vspace{1.5cm}

\begin{center}
\textbf{Abstract}
\end{center}

\noindent The main formal structures of Generalized Quantum Theory are
summarized. Recent progress has sharpened some of the concepts, in
particular the notion of an observable, the action of an observable on
states (putting more emphasis on the role of proposition observables), and
the concept of generalized entanglement. Furthermore, the active role of the
observer in the structure of observables and the partitioning of systems is
emphasized. \newpage

\section{Introduction}

The motivation for the formulation of Generalised Quantum Theory (GQT) \cite%
{ARW,AFR2006}, whose minimal version will be presented here, was the desire
to sharpen the often vague and metaphoric usage of originally quantum
theoretical terms like complementarity and entanglement in fields of
knowledge differing from physics. Starting from the algebraic formulation of
physical quantum theory and discarding from its axioms elements appearing to
be special to physics, a formalism was obtained which is applicable far
beyond the realm of physics and still rich enough to attribute a well
defined formal meaning to the notions of complementarity and entanglement
beyond physics. Meanwhile, quite a few applications demonstrating a partial
structural isomorphy to quantum physics in rather diverse situations have
been worked out in more or less detail (see \cite{ARW}--\cite{Walach2005}).
%,AFR2006,AFR2004,ABFKR,AFR2008, AR2012, Roemer2004,Roemer2006,RW2008,WR2009,%
%Lucadou,Roemer,Roemer2008,LR,Walach2002,Walach2005}).

The formalism of GQT was called ``Weak Quantum Theory'' in the original
publication \cite{ARW} because it arose by weakening the axioms of physical
quantum theory. However, the wider range of applicability of this theory
makes its designation as ``weak'' inappropriate, and for this reason we now
use the term ``Generalized Quantum Theory''. In this short contribution we
will sketch a slight reformulation, simplification and further
generalisation of GQT. This new formulation puts stronger emphasis on the
operationalisable features of GQT and leaves intact all its applications, in
particular the formally precise formulation of complementarity and
entanglement. Details will be given in a more comprehensive publication \cite%
{AFRprep}.

The reader who is not so familiar with the algebraic concepts and the
mathematical formalism of operators, propositions, eigenvectors etc. may
leave aside the mathematical aspects and formulae and stick to the
explanatory remarks.

\section{The formalism of GQT}

GQT takes over from both classical and quantum physics the following four
notions:\\[0.3cm]
\textbf{System}: A system is anything which can be (imagined to be) isolated
from the rest of the world and be subject to an investigation. A system can
be as general as an object or a school of art together with all persons
involved in production and interpretation. Unlike the situation in, e.g.,
classical mechanics the identification of a system is not always a trivial
procedure but sometimes a creative act, in particular in view of the
possibility of entanglement with other systems to be described in section %
\ref{sec4}. In many cases it is possible to define subsystems inside a system%
\\[0.2cm]
\textbf{State}: A system must have the capacity to reside in different
states without losing its identity as a system. One may differentiate
between pure states, which correspond to maximal possible knowledge of the
system and mixed states corresponding to incomplete knowledge. Notice that
the notion of a mixed state contains an epistemic element. We shall see
later that this also applies to some extent to pure states.\\[0.2cm]
\textbf{Observable}: An observable corresponds to any feature of a system,
which can be investigated in a more or less meaningful way. Global
observables pertain to the system as a whole, local observables pertain to
subsystems. (For details concerning the subjective element of partitioning
and identifying subsystems, see section \ref{sec4}.) Observables may, for
instance, correspond to esthetic investigations for systems of art (schools).%
\\[0.2cm]
\textbf{Measurement}: Doing a measurement of an observable $A$ means
performing the investigation which belongs to the observable $A$ and
arriving at a result $a$, which can claim factual validity. What factual
validity means depends on the system: Validity of a measurement result for a
system of physics, internal conviction for self observation, consensus for
groups of human beings. The result of the measurement of $A$ will in general
depend on the state $z$ of the system before measurement but will not be
completely determined by it. \vspace{0.4cm}

To every observable $A$ we associate its \emph{spectrum}, a set Spec\,$A$,
which is just the set of all possible measurement results of $A$.
Immediately after a measurement of an observable $A$ with result a in Spec\,$%
A$, the system will be in an eigenstate $z_a$ of the observable $A$ with
eigenvalue $a$. The eigenstate $z_a$ is a state, for which an immediate
repetition of the measurement of the same observable $A$ will again yield
the same result $a$ with certainty, and after this repeated measurement the
system will still be in the same state $z_a$. This property which is also
crucial in quantum physics justifies the terminology ``eigenstate of an
observable $A$'' for $z_a$ and ``eigenvalue'' for the result $a$. (We should
notice that this does not correspond to a mathematical eigenvalue equation,
however, the relation to the mathematical notion of eigenstate in quantum
physics will be given in Sect.\ \ref{sec3}.) We emphasize that this is an
idealized description of a measurement process.

Even in quantum physics, the attempts to describe the physical measuring
process entirely within its own formalism have not been fully successful so
far \cite{Bell}. GQT deals with systems, whose physical character is not
focal and often unessential. Generalised ``measurement'' in GQT is
predominantly a cognitive process, though, of course, with a physical
substrate.

Two observables $A$ and $B$ are called \emph{complementary}, if the
corresponding measurements are not interchangeable. This means that the
state of the system depends on the order in which the measurement results,
say $a$ and $b$, were obtained. If the last measurement was a measurement of
$A$, the system will end up in an eigenstate $z_a$ of $A$, and if the last
measurement was a measurement of $B$, an eigenstate $z_b$ will result
eventually. For complementary observables $A$ and $B$ there will be at least
some eigenvalue, say $a$, of one of the observables for which no common
eigenstate $z_{ab}$ of both observables exists. This means that it is not
generally possible to ascribe sharp values to the complementary observables $%
A$ and $B$, although both of them are equally important for the description
of the system. This is the essence of quantum theoretical complementarity
which is well defined also for GQT. Sometimes, a stronger notion of
complementarity is employed, which is motivated by the quantum mechanical
complementarity between the position observable $Q$ and the momentum
observable $P$. Here, any definite value of one of these observables means
complete uncertainty of the measurement value of the other observable. This
more restrictive notion of complementarity can be taken over to a large
extent to GQT: We call the observables $A$ and $B$ \emph{strongly
complementary}, if there exists no common eigenstate for both of them.
Strongly complementary observables are, of course, complementary.

Non complementary observables, for which the order of measurement does not
matter, are called \emph{compatible}. After the measurement of compatible
observables $A$ and $B$ with results $a$ and $b$, the system will be in the
same common eigenstate $z_{ab}$ of $A$ and $B$ irrespective of the order in
which the measurements were performed.

\emph{Propositions} are special observables corresponding to ``yes-no''
questions to the system (for the general formalism of quantum mechanics and
propositions, see any text book on quantum theory, e.g. \cite{Thirring}).
Thus the spectrum Spec\,$P$ of a proposition $P$ is contained in the set $\{%
\mathrm{yes, no}\}$. We define trivial propositions \textbf{1} and \textbf{0}
such that \textbf{1} is always true (the result is always ``yes'') and
\textbf{0} is never true. Furthermore, to every proposition $P$ we associate
is negation $\neg P$, which returns the measurement value ``yes'' if $P$
returns ``no'' and returns ``no'' if $P$ returns ``yes''. Evidently, $%
\neg(\neg P) = P$, $P$ and $\neg P$ are compatible, $\neg \mathbf{1}=
\mathbf{0}$. Furthermore, \textbf{1} and \textbf{0} are compatible with all
observables.

For \emph{compatible} (and in fact only for compatible) propositions $P$ and
$Q$ we define a \emph{conjunction} $P\,\mathrm{AND}\,Q$, which gives the
measurement value ``yes'' if and only if both $P$ and $Q$ give ``yes'' and
an \emph{adjunction} $P\,\mathrm{OR}\,Q = \neg(\neg P\,\mathrm{AND}\,\neg Q)$%
, which yields ``yes'' if and only if $P$ or $Q$ yield ``yes''. $P$ and $Q$
are compatible with $P\,\mathrm{AND}\,Q$ and $P\,\mathrm{OR}\,Q$. Simple
identities like
\[
\mathbf{1}\,\mathrm{AND}\,P = P ~~~\mathrm{and} ~~~  P\,\mathrm{AND}\,\neg P
= \mathbf{0}
\]
hold. For details see the original reference \cite{ARW}.

In physical quantum theory states are represented by non-negative hermitean
density matrices $z \neq o$ with $z \geq 0$ and $z^+ = z$. A state $z$ then
associates to an observable $A$ its expectation value $z(A) = \mathrm{tr}%
\,(zA)/\mathrm{tr}\,(z)$. (So, for a number $c \geq 0$, $z$ and $cz$ define
the same expectation value function.) Observables in physical quantum theory
can be identified with mappings, which to every state $z$ associate another
state $A(z) = AzA$. However, $A(z)$ may be the zero matrix, and the price be
paid for defining an action of observables on states is the necessity to
admit an additional improper zero state $o$.

In GQT, states are in general not given by density matrices but one can
still define an action of proposition observables on states after admitting
an improper zero state $o$. Using the structure obtained so far, one defines
for any proposition $P$

\begin{itemize}
\item[-] $P(o) = o$, and for $z \neq o$

\item[-] $P(z) = o$, if the measurement result ``yes'' for $P$ is impossible
in the state $z$

\item[-] $P(z) =$ the state obtained after a measurement of $P$ with result
``yes'', if the result ``yes'' is possible in the state $z$.
\end{itemize}

This definition imitates the action of propositions on states in quantum
physics. The action of propositions $P$ on states employs measurements of $P$
and is operationally well defined.

We immediately see
\[
\mathbf{0}(z) = 0 ~~ , ~~~ \mathbf{1}(z) = z
\]

Moreover, for (compatible or complementary) propositions $P$ and $Q$ we can
define a composition $PQ$ as mappings on states by
\[
PQ(z) = P(Q(z))
\]
and we have
\begin{eqnarray*}
& & PP(z) = P(z)\, ,~ \mathrm{i.e.}~~ PP = P \\
& & P\, \neg P = \neg P\, P = 0 \\
& & \mathbf{0}P = P\mathbf{0} = \mathbf{0}~,~~ \mathbf{1}P = P\mathbf{1} = P
\end{eqnarray*}
A proposition $P$ can be identified with the pair of mappings on states
associated to $P$ and $\neg P$, and two propositions $P$ and $Q$ are
compatible, if and only if all the propositions $P$, $\neg P$, $Q$, $\neg Q$
commute as mappings on states, i.\,e., if and only if
\[
PQ = QP~,~ P\, \neg Q = \neg Q\, P~,~ \neg P\,Q = Q\, \neg P~,~  \neg P\,
\neg Q = \neg Q \, \neg P \, .
\]
For \emph{compatible} propositions $P$ and $Q$
\[
P~\mathrm{AND}~ Q = PQ = QP
\]
holds.

Quite generally, observables can be reduced to families of propositions in
the following way: For every $a$ in Spec\,$A$ let $A_a$ denote the
proposition corresponding to the assertion that the value of $A$ is $a$. In
particular, for a proposition $P$ we have
\[
P_{\mathrm{yes}} = P ~~~\mathrm{and}~~~ P_{\mathrm{no}} = \neg P \, .
\]
Evidently
\[
A_a A_{a^\prime} = \mathbf{0} ~~ \mathrm{for}~ a \neq a^\prime
\]
and
\[
\bigcup_{a \in \mathrm{Spec}\,A} A_a = \mathbf{1} \, .
\]
This simply means that the propositions $A_a$ are mutually exclusive and
that after a measurement exactly one of them has to be true.

In fact an alternative way to build up the formalism of GQT would be to
start out from propositions and to define observables $A$ as pairs $(\mathrm{%
Spec}\, A, (A_a)_{a \in \mathrm{Spec}\,A})$ consisting of the set Spec\,$A$
and a family of propositions $A_a$ with the properties defined by the last
two formulae.

The observable $A$ and the propositions $A_a$ are compatible and two
observables $A$ and $B$ are compatible if and only if all the propositions $%
A_a$ and $B_b$ are compatible.

In the original formulation of GQT, like in physical quantum theory, the
existence of an action of observables on states was postulated. Even in
physical quantum theory this action is operationally not easily accessible.
In our new formulation we restrict ourselves to the introduction of an
operationally better defined action of propositions on states. This
simplification and further weakening of the axioms of GQT has no affect on
the definition of a generalized notion of complementarity and, as we shall
see, entanglement.

In concrete applications of GQT, however, it may be advantageous to go
beyond the minimal formal scheme of GQT and use an action of all observables
on states. This is for instance the case for the application of GQT to
bistable perception quoted in ref.\ \cite{AFR2004}.

\section{Comments on some aspects of GQT}

\label{sec3}

Comparing physical quantum theory with the (original or new formulation of)
GQT one readily finds the following essential differences:

\begin{itemize}
\item In GQT there is no quantity like Planck's constant controlling the
degree of complementarity of observables. Thus, contrary to physical quantum
theory, where quantum effects are essentially restricted to the microscopic
regime, macrosopic quantum like effects in GQT are to be expected.

\item At least in its minimal version described here, GQT contains no
reference to time or dynamics.

\item In its minimal version GQT does not ascribe quantified probabilities
to the outcomes of measurements of an observable $A$ in a given state $z$.
Indeed, to give just one example, for esthetic observables quantified
probabilities seem to be inappropriate from the outset. What rather remains
are modal logical qualifications like ``impossible'', ``possible'' and
``certain''. Related to the absence of quantified observables, the set of
states in GQT is in general not modeled by a linear Hilbert space. Moreover,
no addition of observables (operationally difficult to access even in
quantum physics) is defined in GQT.

\item In quantum physics where states are defined by means of a linear
Hilbert space observables can be identified with linear maps of the Hilbert
space into itself. Furthermore, an eigenstate of an observable $A$ is a
state which is reproduced by the linear action of observables on states. In
GQT only proposition observables act on states, as mentioned above.
Moreover, linearity of this action cannot be defined. The salient feature of
an eigenstate in quantum mechanics is that further measurements of the same
observable without intervening disturbations will reproduce it. This
decisive and operationally well-defined feature is assumed in our definition
of an eigenstate in GQT.

\item Related to this, GQT in its minimal form provides no basis for the
derivation of inequalities of Bell's type for measurement probabilities,
which allow for the conclusion, that the indeterminacies of measurement
values are of an intrinsic nature rather than a lack of knowledge. It is an
open question whether the construction of states analogous to GHZ-states in
quantum physics \cite{GHZ}, which allow for a decision in favour of ontic
indeterminacies by one single measurement, is possible in GQT. As such, GQT
is a phenomenological theory and leaves the question for the ontic or
epistemic nature of indeterminacies open. The answer may depend on the
concrete system considered. After all, physical quantum with ontic
indeterminacies is a special case of GQT. In many applications of GQT
indeterminacies may be epistemic and due to incomplete knowledge of the full
state or uncontrollable perturbations by outside influences or by the
process of measurement. Notice that complementarity in the sense of GQT may
even occur in coarse grained classical dynamical systems \cite%
{Graben2006,Graben2008}.

\item For some applications (see, e.g., \cite{AFR2004,ABFKR,AFR2008}) one
may want to enrich the above described minimal scheme of GQT by adding
further structure, e.g., an underlying Hilbert space structure for the
states.
\end{itemize}

We should stress here that for very general systems like schools of art,
observables are not so directly given by the system and read off from it
like location and velocity in a mechanical system. On the contrary, as
already suggested by the name of an ``observable'', the identification of an
observable may be a highly creative act of the observer, which will be
essentially determined by his horizon of questions and expectations. This
marks a decidedly epistemic trait of the notion of observables in GQT even
more than in quantum physics. Moreover, the horizon of the observer will
change, not the least as a result of his previous observations adding to the
open and dynamical character of the set of observables.

\section{Partitions and Entanglement}

\label{sec4}

What has just been said about observables also applies to \emph{partitioning}
a system into subsystems. In fact, partitioning is achieved by means of
\emph{partition observables} whose different values differentiate between
the subsystems. In general, subsystems do not preexist in a na\"\i{}ve way
but are in a sense created in the constitutive act of their identification.

Different partitions may be compatible or complementary. The physical
position observable $Q$ is a privileged example of a partition observable
defining a partition into spacially separated subsystems. The range of
applicability of this partition observables is largely coincident with the
range of physics (physics as the realm of ``res extensae'').

The first partition prior to every further one and prerequisite for any act
of cognition is the \emph{epistemic split} into observer and observed
system: After all, any cognition of the kind we have access to is the
cognition of someone about something. One might speculate about the
existence of splittings which are complementary to any epistemic split.

In quantum physics, the epistemic split is known under the name of \emph{%
Heisenberg split}. It can be moved and shifted but it is inevitable for any
measurement. In the quantum theory of physical measurement one can observe
that the stochasticity of measurement results enters as a result of the
Heisenberg split and subsequent projection onto either the subsystem of the
observer or the observed object. There is a symmetry in the probabilities of
measured values with respect to projection onto measured system and
measuring device. One might ask oneself, to what extent this symmetry could
be generalized in GQT.

The genuinely quantum theoretical phenomenon of entanglement can and in
general will show up also in GQT if the following conditions are fulfilled:

\begin{enumerate}
\item A system is given for which subsystems can be identified. Entanglement
phenomena will be best visible if the subsystems are sufficiently separated.
Local observables pertaining to different subsystems are compatible.

\item There is a global observable of the total system, which is
complementary to the local observables of the subsystems.

\item Given these two prerequisites, a system will be in an \emph{entangled
state} if this state is an eigenstate of the above mentioned global
observable and not an eigenstate of the local observables. (In GQT this is
the definition of an entangled state, while in quantum theory this can be
shown to be equivalent to the usual definition of an entangled state as not
being separable; see, e.g., \cite{Nielsen}, which in general is a good
reference for all questions concerning entanglement in quantum physics.)
\end{enumerate}

Given these conditions, the measured values of the local observables will be
uncertain because of the complementarity of the global and the local
observables. However, so-called \emph{entanglement correlations} will be
observed between the measured values of the local observables (see figure
1). These correlations are non local and instantaneous. Einstein, trying to
argue for an incompleteness of quantum mechanics, spoke about ``spooky
interactions'' in this connection. Entanglement correlations have been
observed beyond any doubt in quantum physics (see \cite{Nielsen} and
references therein). Entanglement correlations are not due to causal
interactions between the subsystems.

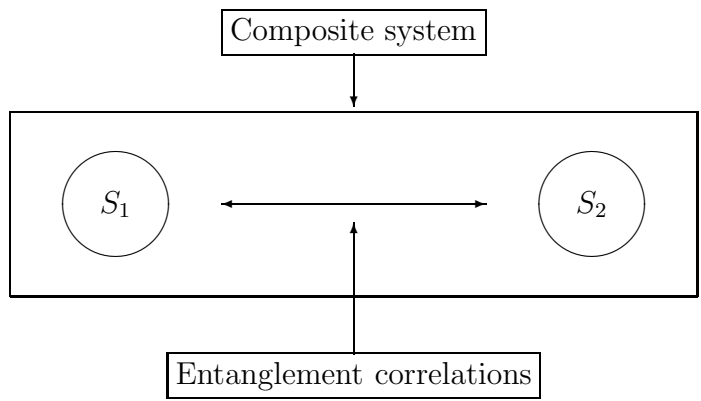
\begin{figure}[tbh]
\begin{picture}(300,150)(0,0)
\put(150,140){\makebox(0,0){\fbox{Composite system}}}
\put(150,132){\vector(0,-1){20}}
\put(20,40){\line(1,0){260}}
\put(20,40){\line(0,1){70}}
\put(20,110){\line(1,0){260}}
\put(280,40){\line(0,1){70}}
\put(60,75){\circle{40}}
\put(240,75){\circle{40}}
\put(150,18){\vector(0,1){50}}
\put(100,75){\vector(1,0){100}}
\put(110,75){\vector(-1,0){10}}
\put(60,75){\makebox(0,0){$S_1$}}
\put(240,75){\makebox(0,0){$S_2$}}
\put(150,10){\makebox(0,0){\fbox{Entanglement correlations}}}
\end{picture}
\caption{Entanglement}
\end{figure}

Such correlations without interactions are a witness of the holistic
character of composite quantum systems: In general, the states of the
subsystems do not determine the state of the total system. Vice versa, the
holistic state of the total system does not determine the measured values of
local observables pertaining to the subsystems. The holistic character of
the total quantum state resides in entanglement correlations between the
subsystems.

It is not difficult to show that in quantum physics entanglement
correlations cannot be used for signal transmission between different
subsystems. This last statement is sometimes referred to as Eberhards
theorem \cite{Eberhard}. For a simple explanation see \cite{Lucadou}. This
must also hold in GQT in order to prevent bizarre intervention paradoxes.
One may even turn the argument around and state that, whenever correlations
between subsystems can be used for signal transfer, they must be of causal
nature and entanglement must be absent or at least not dominant.

For completeness, we should mention here, that a distinction between local
and global observables is also possible for temporal separations (see, for
instance \cite{AFR2008}).

As already mentioned above, in quantum physics inequalities for entanglement
correlations of Bell's type can be employed to show that the indeterminacies
of the measured values of local observables, and, more generally, of any
quantum observable, can be of ontic rather than epistemic nature. Bell\~{O}s
inequalities also allow to differentiate between a genuine entangled state
and a mixture of product states, which will show similar correlations
between subsystems.

In GQT in its minimal form there are no Bell equalities and the above
differentiation may be difficult or impossible. In such situations, GQT
should be considered a phenomenological theory which leaves the question for
the ontic or epistemic character of indeterminacies open.

We already saw that observables in quantum theory, and even more so in GQT,
are not exclusively to be attributed to the observed system but have a
strong reference to the observer. They are to be located on the epistemic
split and hence have a certain epistemic connotation.

This also applies to the notion of a state. In classical mechanics, a pure
state can be interpreted as an entirely ontic entity. In quantum physics and
certainly also in GQT, the notion of a state is not so much related to
``what there is'' in the system but rather to what will be observed. This
gives a definite epistemic colouring even to the predominantly ontic notion
of a pure state, which applies in particular if GQT is interpreted as a
phenomenological theory of systems. \vspace{0.7cm}

\noindent \textbf{Acknowledgement}\\[0.2cm]
Both authors acknowledge numerous discussions with Harald Atmanspacher.

\end{document}